\documentclass[a4paper]{article}

\usepackage{INTERSPEECH2021}
\usepackage{url}
\usepackage{multirow}
\usepackage{pifont}
\newcommand{\cmark}{\ding{51}}%

\usepackage{xcolor}

\title{Deformable CNN and Imbalance-Aware Feature Learning \\for Singing Technique Classification
}

\name{Yuya Yamamoto$^1$, Juhan Nam$^2$,  Hiroko Terasawa$^1$}
\address{
$^1$Doctoral Program in Informatics, University of Tsukuba, Japan\\
$^2$Graduate School of Culture Technology, KAIST, South Korea}
\email{s2130507@s.tsukuba.ac.jp, juhan.nam@kaist.ac.kr, terasawa@slis.tsukuba.ac.jp}

\begin{document}

\maketitle

\begin{abstract}
Singing techniques are used for expressive vocal performances by employing temporal fluctuations of the timbre, the pitch, and other components of the voice.
Their classification is a challenging task, because of mainly two factors:
1) the fluctuations in singing techniques have a wide variety and are
affected by many factors and 2) existing datasets are imbalanced.
To deal with these problems, we developed a novel audio feature learning
method based on deformable convolution with decoupled training of the
feature extractor and the classifier using a class-weighted loss function.
The experimental results show the following: 1) the deformable
convolution improves the classification results, particularly when it is applied to the last two convolutional layers, and 2) both re-training the classifier and weighting the cross-entropy loss function by a smoothed inverse frequency enhance the classification performance.
\end{abstract}

\section{Introduction}

Professional singers express their characteristics and emotions by various singing techniques such as vibrato and breathy voice effects. At the signal level, singing techniques are observed as time--frequency textures, e.g., strong temporal modulation related to harmonics (``vibrato'') and highly noisy components over broad frequency bands (``breathy voice''). Automatic classification of singing techniques is an emerging research topic in singing voice analysis \cite{humphrey2018introduction}.


One of the main problems in this task is extracting features from highly dynamic time--frequency textures of singing techniques.
Convolutional neural networks (CNNs) have been recently used as effective methods to capture audio features for singing technique classification \cite{wilkins2018vocalset, yamamoto2021investigating, o2021zero} as well as similar objectives such as musical playing technique recognition \cite{abesser2019fundamental}.
Although square-shaped kernels, e.g., 3$\times$3 and 5$\times$5, are commonly used in CNNs, it has been shown that customizing the kernel shape improves the classification performance. For example, in a study, oblong-shaped kernels outperformed square-shaped ones for singing technique classification \cite{yamamoto2021investigating}. Similar results have been found in music auto-tagging \cite{pons2016experimenting} and musical instrument classification \cite{takahashi2018instrudive}.
The above findings suggest that more customized kernels may further improve the performance. However, a brute-force search toward the best kernel shape will be burdensome, and thus, a systemic approach is required.

Another critical problem in singing technique classification is data
imbalance, which is mainly attributed to the nature of voice production and musical usage. For example, ``vocal fry'' and ``trillo'' are difficult to produce for a long time, and thus, the lengths of such audio samples tend to be relatively short. In addition, ``belting'' is obtained in only certain musical contexts. Thus, collecting well-balanced samples is problematic.

In this study, we deal with the above two problems by deformable convolution and classifier re-training (cRT) using a class-weighted loss, respectively. Deformable convolution allows the convolution kernel to have a flexible shape \cite{dai2017deformable}. It extends the capability of a CNN by modeling geometric transformation, which can be beneficial in capturing dynamic time--frequency features in singing techniques. cRT decouples the feature extractor and the classifier in training a deep neural network model. It was reported as a simple yet powerful method when the class distribution of the training data has a long tail \cite{Kang2020Decoupling}.


The contributions of this study are as follows: 1) We investigate
different setups of deformable convolution and show that it improves the singing technique classification performance. 2) We show the effectiveness of cRT for an imbalanced dataset, comparing with joint training of the feature extractor and the classifier. 3) Finally, we present that smoothed weighting the loss function further enhances the effect of decoupled training.

\begin{figure*}[!htbp]
  \centering
  \includegraphics[width=0.92\textwidth]{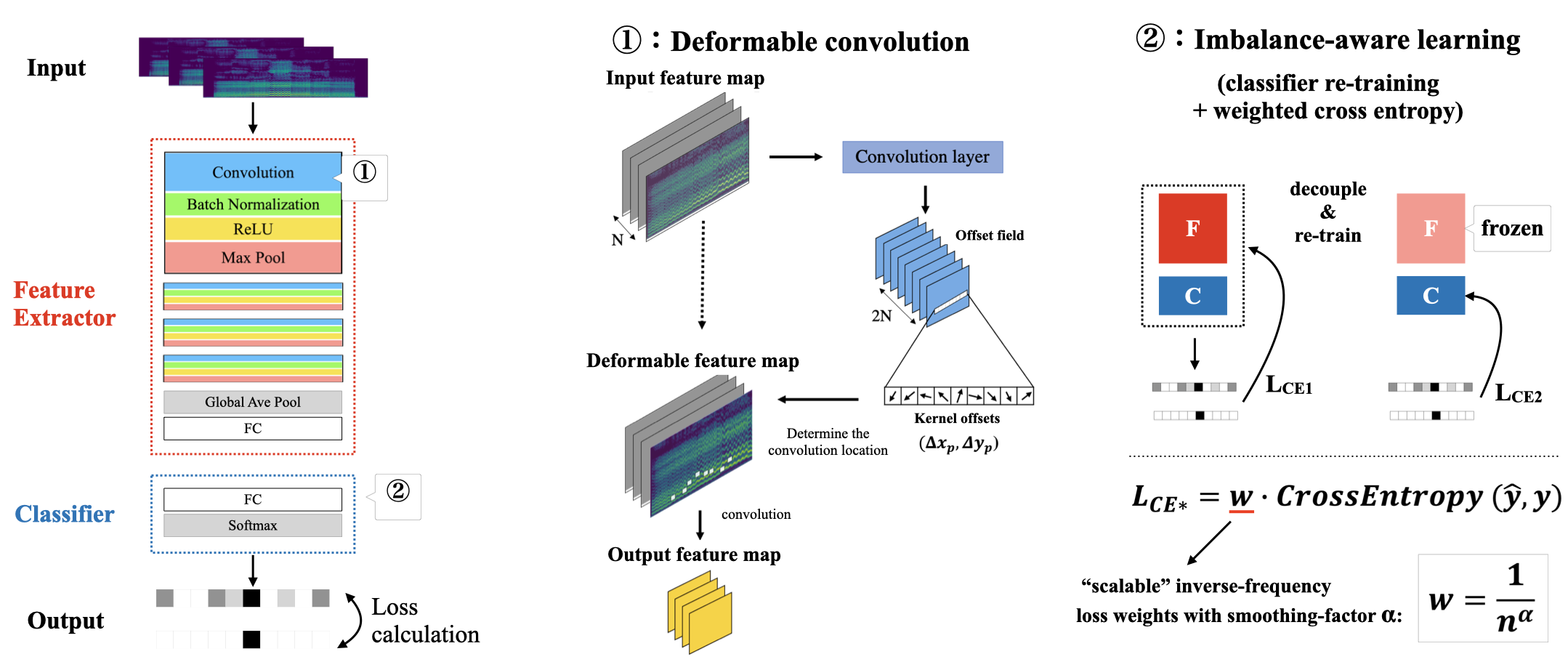}
  \caption{Overview of the proposed method for singing technique classification 
}
  \label{fig:overview}
\end{figure*}

\section{Related Work}

\subsection{Deformable Convolution}
Deformable convolution was introduced for image processing to enhance the transformation modeling capability of a CNN \cite{dai2017deformable, zhu2019deformable}. 
It allows CNN models to only focus on what they are interested in and makes the output feature maps more representative. 
It has been effective in several tasks involving variations in the temporal context, such as action recognition \cite{lei2018temporal}, sign language recognition \cite{papadimitriou2020multimodal}, and video captioning \cite{chen2019temporal}. 
In the audio domain, deformable convolution has been employed in speaker verification  \cite{zhang2020speaker} and speech recognition \cite{an21_interspeech}, to deal with the variable temporal dynamics of speech. In this study, we apply deformable convolution to singing technique classification.

\subsection{Data Imbalance}
Data imbalance is a common issue in classification tasks. There are two well-known approaches for solving this problem: sampling and cost-sensitive learning \cite{sun2009classification}.
Sampling manipulates the class representations in an original dataset by either over-sampling the minority classes (over-sampling) or under-sampling the majority classes (under-sampling). In the context of deep learning, neither over-sampling nor under-sampling is efficient; over-sampling decelerates the training and may cause overfitting, whereas under-sampling may discard informative majority examples \cite{cui2019class}.
Cost-sensitive learning is a type of learning that considers the
misclassification costs. 
A simple approach of cost-sensitive learning is reweighting the loss function using inverse class frequency values \cite{huang2016learning}. However, this strategy may perform poorly when applied to real-world and large-scale datasets. Comparatively,
``smoothed'' weighting (e.g., square root of the class frequency values \cite{mahajan2018exploring} and heuristically determined exponent values \cite{xue2021mt5}) is known to be more effective.
Data imbalance was recently addressed in a study by decoupling the feature extractor and the classifier during training of deep neural networks \cite{Kang2020Decoupling}.
Empirical experiments showed that the data imbalance problem affects
learning classifier decision boundaries, instead of learning feature
representations. In this study, we investigate the smoothed weighting and
decoupling of the feature extractor and the classifier.

\section{Method}
Figure \ref{fig:overview} shows an overview of our proposed method, and this section describes the details of each of its parts.

\subsection{Deformable Convolution}
Deformable convolution (DC) facilitates trainable offset parameters of each kernel to deform the convolutional kernel grid.
Deformable convolution consists of the following steps:
1) obtain the offset field, 2) output deformable feature maps by the offsets, 3) perform regular convolution on the deformable feature maps. 
The middle of Figure \ref{fig:overview} illustrates the operation of deformable convolution.

\begin{enumerate}
    \item The offset field is obtained by applying a convolutional layer over the input feature map with channel dimension $N$.
    The offset field has the same spatial resolution as the input feature map, and the channel dimension is $2N$. Horizontal offset $\Delta x$ and vertical offset $\Delta y$ correspond to each point of the input feature map. 
    
    
    \item Because the offset parameters ($\Delta x$ and $\Delta y$) are typically fractional, the values of offset location are interpolated around the value of closest four points by bilinear interpolation \cite{dai2017deformable}.
  
    \item The output feature maps are obtained by operating a regular convolution using the deformable feature maps.
    For each location $p_0$ on the input feature map $x$, and the output feature map $y$,
    $$y(p_0) = \sum_{m \in R}^{M}{w(p_0) \cdot x(p_0 + p_m + \Delta p_m)}$$
    $$ p_m = (\Delta x_{p_m}, \Delta y_{p_m})$$
    where $w$, $R$, and $p_m$ denote the weight of the sampled values, kernel grid with size $M$, and interpolated offset value, respectively.
\end{enumerate}

 We choose a four-oblong-shaped convolution layer CNN with a multi-resolution spectrogram input, which was the best performing model in \cite{yamamoto2021investigating}, for the base architecture.
 The model consists of four convolutional blocks, a global average pooling layer \cite{lin2013network} \footnote{In the original study, a flatten layer was used in the top part of the feature extractor. However, in singing technique classification, we confirmed that the global average pooling layer generally outperforms the flatten layer.}, and two fully connected layers.


Note that although its kernel shapes are unidirectional  (i.e., $(Vertical, Horizontal) =  \{(4\times1) , (16 \times 1), (1 \times 4), (1 \times 16)\}$ ), we consider both vertical and horizontal offsets as same, similar to conventional studies \cite{dai2017deformable,zhu2019deformable, zhang2020speaker}, to preserve flexibility.

\subsection{Weighting Loss Function}
We apply a smoothed weighting to the cross-entropy loss function during training, to deal with the data imbalance problem.  
\begin{equation}
\mathrm{L}(x, y) = - W \log{\frac{\exp(x_{n,y_n})}{\sum_{c=1}^C \exp(x_{n,c})}}
\end{equation}
where $x$ is the input, $y$ is the target, and $W$ is the weight of the loss function.
We determine the loss weight of each class $w_c$ by the power of the inverse frequency of the training sample as follows: 
\begin{equation} \label{eq:alpha}
    w_c = \frac{1}{(n_c) ^ \alpha}
\end{equation}
where $n_c$ is the number of training samples in class $c$, and
$\alpha$ is the smoothing factor, controlling smoothing of the loss weights.
Note that $\alpha$ = 0 corresponds to the value of 1 (i.e., no weighting) and $\alpha$ = 1 corresponds to a reciprocal number (i.e., weighting by the inverse class frequency).

\subsection{Decoupling Feature Extractor and Classifier}
We also investigate decoupling the feature extractor and the classifier from the CNN model, following the method of Kang et al. \cite{Kang2020Decoupling}. They proposed two different approaches of the decoupled training method: cRT and normalizing the weights of the classifier by its own norms scaled by a hyperparameter ($\tau$-normalized classifier), called learnable weight scaling, and showed that both outperformed joint training of the classification model. We choose to employ cRT, which was reported as a simple but effective training strategy for an imbalanced dataset. First, the layers of the model are divided into two parts---the feature extractor and the classifier---between the first and second fully connected layers.
In the training stage, first the model is trained regularly and subsequently the classifier is re-trained after fixing the weights of the feature extractor part. The right panel of Figure \ref{fig:overview} illustrates the training strategy.

\section{Experiments}
\subsection{Dataset}
\begin{figure}[!t]
  \centering
  \includegraphics[width=1.02\columnwidth]{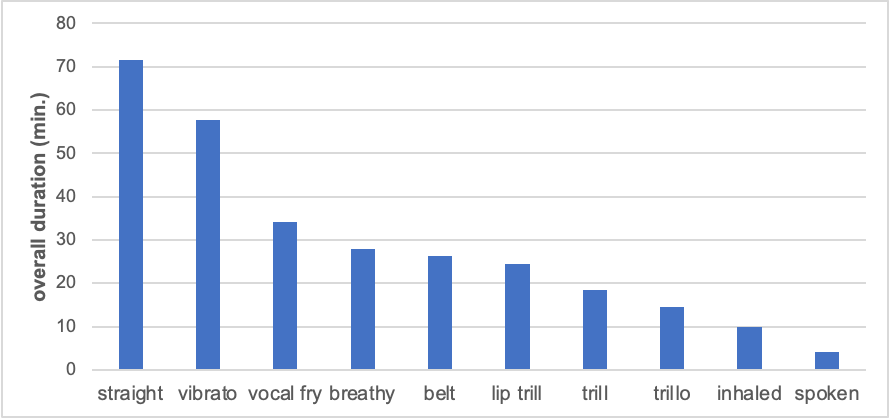}
  \caption{Audio length distribution of singing techniques in VocalSet \cite{wilkins2018vocalset}. }
  \label{fig:dist_vocalset}
\end{figure}

 We use VocalSet \cite{wilkins2018vocalset}, which is the only publicly available dataset for studies on singing techniques. 
 The dataset contains singing voices of 20 different professional singers (9 female and 11 male) performing 17 different singing techniques in various contexts, such as arpeggio, scale, and long tones. 
 For the classification experiments, we select the samples corresponding to ten different singing techniques (``belt,'' ``breathy,'' ``inhaled singing,'' ``lip trill,'' ``spoken excerpt,'' ``straight tone,'' ``trill,'' ``trillo,'' ``vibrato,'' and ``vocal fry''). Figure \ref{fig:dist_vocalset} shows the total length of each singing technique. The distribution of the dataset has a long-tail shape, i.e., it is imbalanced.

 During the learning process, we split the dataset into a training set of 15 singers and a test set of 5 singers \footnote{The train and test split is officially provided. Refer to the file, ``train\_singers\_technique.txt'' in Version 1.2 
\\ https://zenodo.org/record/1442513\#.YjjqlJrP3a4}.
 Subsequently, we segment the audio signals in each file into 3-second audio clips and nonoverlapping parts at a sample rate of 44.1 kHz. We evaluate each model using five metrics: macro-F1 score (F1), balanced accuracy (B-Acc.), accuracy (Acc.), top-2 accuracy, and top-3 accuracy.

\subsection{Model \label{Sec:Model}}
We set up four types of deformable convolution model (DC) with weighting and two models without deformbale convolution (w/o DC) with or without weighting. As a result, we compare six conditions in total. For all of these six conditions, the model input and structure are common as follows. The model input is multi-resolution spectrogram (i.e., stacking three spectrograms with different time-frequency resolutions along the channel dimension.) We obtain them by short-time Fourier Transform (STFT), and each spectrogram is obtained by the three window sizes of (2048, 1024, 512 samples) with the same hop length 512 samples and the STFT length 2048 samples with zero-padding. 
We employed a four-oblong-shaped convolution layer CNN \cite{yamamoto2021investigating} for the model structure. Each convolution block consists of a convolution layer (Conv), a batch normalization layer, a Rectified Linear Unit (ReLU), a max pooling (MP) layer, and a dropout of 0.3. They are followed by a global average pooling (Global AP) layer and two fully-connected layers (FC). 
We trained our model using the Adam optimizer with a learning rate of 1e-4 and a batch size of 64.

The four DC conditions are denoted as \emph{All}, \emph{Early}, \emph{Late}, and \emph{Last} and their components are listed in Table \ref{tab:stft_cnn}. DC is applied to different layers. All DC models are trained with the weighted loss-function. For the non-DC models, we considered two w/o DC conditions with or without weighting, they are referred to as \emph{w/o DC weighted} and \emph{w/o DC plain}. 

\begin{table}[!t]
\begin{center}
\caption{Configuration of the model. The Four DC conditions differ in the arrangement of DC application. The check mark represents DC application to the corresponding layer.}
\begin{tabular}{l|l|p{0.015\textwidth}p{0.02\textwidth}p{0.02\textwidth}p{0.02\textwidth}}
\hline
\multirow{2}{*}{Layer Configuration} & \multirow{2}{*}{Ch} & \multicolumn{4}{c}{Deformable Conv} \\ 
\cline{3-6}
& & All & Early & Late & Last \\
\hline\hline
Conv($4\times1$), MP($4\times4$) & 32 & \cmark & \cmark & &\\
Conv($16 \times 1$), MP($4 \times 4$) & 64 & \cmark & \cmark & &\\
Conv($1 \times 4$), MP($3 \times 3$) & 128 & \cmark & & \cmark&\\
Conv($1 \times 16$), MP($2 \times 2$) & 128 & \cmark & & \cmark & \cmark\\
 Global AP & 128 & \multicolumn{4}{c}{--}\\
 FC (Feature) & 30 & \multicolumn{4}{c}{--}\\
 FC (Softmax) & 10 & \multicolumn{4}{c}{--}\\
\hline
\end{tabular}
\vspace{-3mm}
\label{tab:stft_cnn}
\end{center}
\end{table}

\subsection{Experiment 1: Effect of Deformable Convolution}

We investigate the effect of deformable convolution by replacing standard convolution layers of the model with deformable convolution layers. We tested the six conditions as described in Section \ref{Sec:Model}. 
As baselines, we use one-dimensional CNN (1DCNN) \cite{wilkins2018vocalset} and oblong-CNN feature learning with a random forest classifier (Oblong) \cite{yamamoto2021investigating}. We re-implement the models to investigate the effect of weighting the loss function. For both 1DCNN and Oblong, we tested both weighted and plain (without weighting) conditions. The number of parameter of each conditions are as follows; w/o DC: 337.5k, All: 463.3k, Early: 362.2k, Late: 438.7k, and Last: 435.7k, respectively.


\subsection{Experiment 2: Comparative analysis of training strategy and $\alpha$}

We compare three training strategies with a set of smoothing factors $\alpha$ (0, 0.2, 0.5, and 1) in Eq. \ref{eq:alpha} seeking the best DC setup. 

\begin{itemize}
    \item \textbf{Joint training}: without classifier retraining.
    \item \textbf{cRT-WFC}: weights are applied during \textbf{both} \textbf{feature representation} training and \textbf{cRT} phases.
    \item \textbf{cRT-WC}: weights are applied \textbf{only} during the \textbf{cRT} phase. (i.e., weights are not applied in the feature representation training phase)
\end{itemize}

These training strategies were tested upon the \emph{Late} model because it was the best model in experiment 1 as described in Section \ref{sec:resDC}. For reasonable comparison, the sum of the number of training epochs is set equal in all conditions. We set 200 epochs for the entire training time. For all cRT-based methods, we assign 100 epochs for the joint training of the feature extractor and the classifier, and the remaining 100 epochs for the cRT.


\section{Results and Discussions}

\subsection{Effect of Deformable Convolution \label{sec:resDC}}
The results of experiment 1 are listed in Table \ref{tab:results_deform}. They show that DC models significantly improve the classification performance compared to w/o DC models. Among the four DC setups, the \emph{Late} model achieves the best. This agrees with the results from previous works that applying DC to several late convolution layers is effective \cite{zhu2019deformable, an21_interspeech}. 
Compared to the \emph{Last} model where DC is applied only to the last convolution layer, the accuracy of the \emph{Late} model becomes much higher. Class-wise accuracy may explain this gap: With the \emph{Late} model we observed large accuracy increments on the discrimination of ``lip trill'' and ``vocal fry,'' which have fine temporal modulation in amplitude, frequency, and breathiness. 

This indicates that the small kernel size of the 3rd DC layer plays an important role when the dynamic offset adapts the fine modulation of singing voice. 
The baseline model with Oblong kernel-shapes achieves higher accuracy than the model without DC, as it uses a random forest classifier on a similar configuration of CNN feature extractor. 
However, the \emph{Late} model extracts the features more effectively with DC and outperforms the baseline model.   



\begin{table}[!t]
    \centering
    \caption{The results of experiment 1. 
    }
    \scalebox{0.85}{
    \begin{tabular}{l|ccccc}
    \hline
    \begin{tabular}{l}
        Models
    \end{tabular}
         & F1 & Acc. & B-Acc. & Top-2 & Top-3 \\ \hline \hline
         \hline
         1DCNN \cite{wilkins2018vocalset} plain & 0.488 & 0.584 & 0.484 & 0.764 & 0.863 \\ 
         1DCNN \cite{wilkins2018vocalset} weighted & 0.306 & 0.439 & 0.352 & 0.643 & 0.753 \\ 
         Oblong \cite{yamamoto2021investigating} plain & 0.540 & 0.600 & 0.597 & 0.757 & 0.838 \\
         Oblong \cite{yamamoto2021investigating} weighted & 0.548 & 0.590 & 0.613 & 0.759 & 0.852 \\
        
        \hline
        
        w/o DC plain & 0.404 & 0.492 & 0.472 & 0.686 & 0.805 \\
        w/o DC weighted & 0.513 & 0.554 & 0.575 & 0.743 & 0.858 \\  
        All (1,2,3,4) & 0.553 & 0.604 & 0.59 & 0.799 & \textbf{0.896}\\ 
        Early (1,2) & 0.554 & 0.593 & 0.598 & 0.776 & 0.862 \\ 
        Late (3,4) & \textbf{0.582} & \textbf{0.623} & \textbf{0.641} &\textbf{0.806} & 0.894 \\ 
        Last (4) & 0.517 & 0.572 & 0.607 & 0.764 & 0.846 \\ \hline
    \end{tabular}
    }
    \label{tab:results_deform}
\end{table}


\begin{table}[!t]
    \centering
    \caption{The results of comparison between joint-training, cRT-WC and cRT-WFC, under $\alpha = 0.2$.}
    \begin{tabular}{l|ccccc}
    \hline
        Methods & F1 & Acc. & B-Acc. & Top-2 & Top-3 \\ \hline \hline
        Joint-training & 0.559 & 0.610 & 0.635 & 0.774 & 0.874 \\ \hline
        cRT-WFC & 0.582 & 0.623 & 0.641 & 0.806 & \textbf{0.894} \\ 
        cRT-WC & \textbf{0.620} & \textbf{0.656} & \textbf{0.655} & \textbf{0.815} & 0.887 \\ \hline
    \end{tabular}
    \label{tab:joint_vs_crt}
\end{table}

\subsection{Effect of cRT}
Table \ref{tab:joint_vs_crt} shows the results for the training strategies comparison, summarizing the output with the smoothing factor $\alpha$ = 0.2 (as discussed in Section \ref{Sec:alpha}.) Both cRT methods outperform the joint-training method. Between two cRT methods, cRT-WC significantly improves the classification performance. This suggests that the weighting loss-function is only effective in cRT and so it is better to apply the weighting only during the re-training phase. A similar result was also reported in \cite{Kang2020Decoupling}.   



\begin{figure}[!t]
  \centering
  \includegraphics[width=\columnwidth]{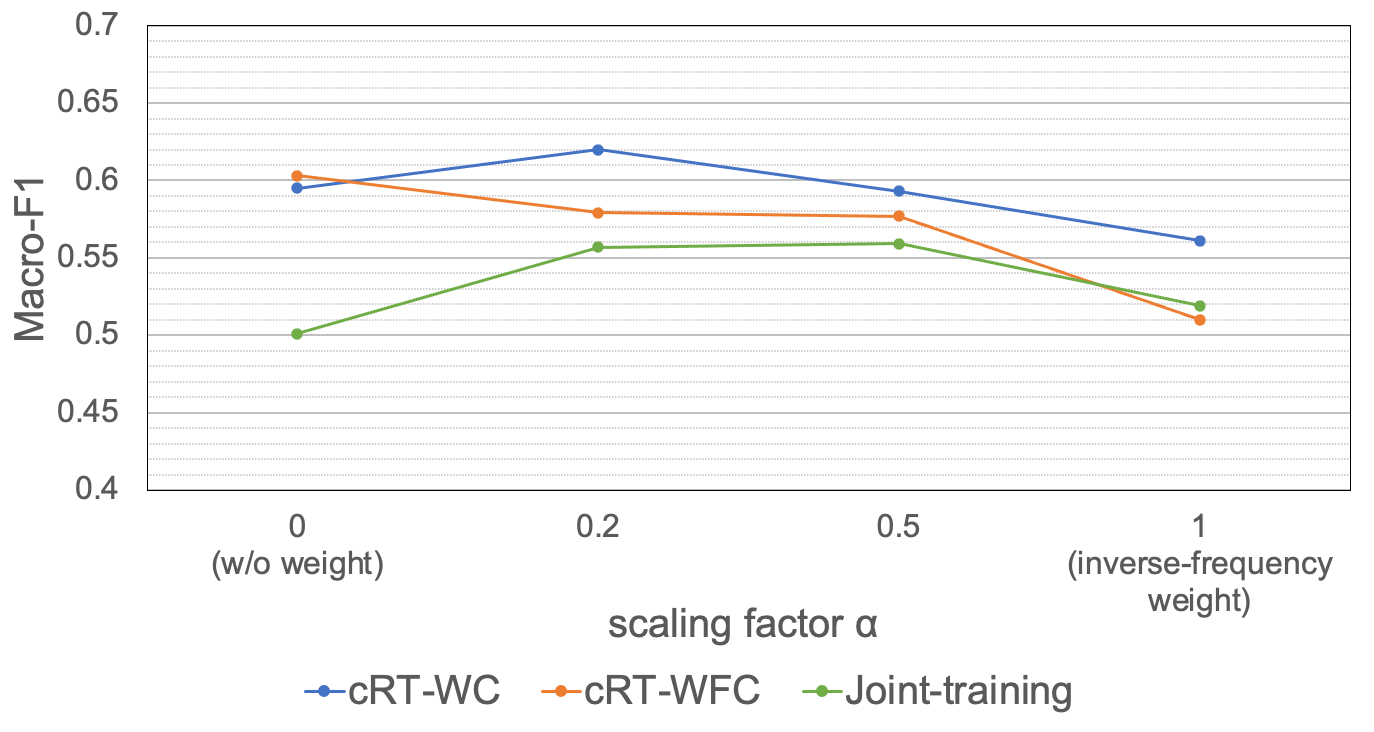}
  \caption{Macro F1 score for cRT-WC, cRT-WFC, and joint-training respectively with four different $\alpha$ values.}
  \label{fig:ablation_alpha}
\end{figure}

\subsection{Effect of Smoothing Factor $\alpha$ \label{Sec:alpha}}
We conducted experiment 2 with four different values of the smoothing factor $\alpha$; 0, 0.2, 0.5, and 1. Figure \ref{fig:ablation_alpha} plots Macro-F1 over the smoothing factor. The best performing condition is cRT-WC with an $\alpha$ value of 0.2. As $\alpha$ increases, the performance keeps decreasing in all three conditions and reaches the worst accuracy at an $\alpha$ value of 1 (i.e., inverse-frequency weight).      


Increasing $\alpha$ has the expected effect of improving performance of minority classes while hurting majority classes. However, when we vary $\alpha$ from 0.2 to 1.  the class-wise F1 scores decreased for both minority (e.g., ``inhaled'' 0.293 $\rightarrow$ 0.268, ``trill'' 0.544 $\rightarrow$ 0.495) and majority (e.g., ``straight'' 0.69 $\rightarrow$ 0.645, ``vibrato'' 0.648 $\rightarrow$ 0.623.) It corresponds to the result of conventional works \cite{mahajan2018exploring, cui2019class} that inverse frequency weight decreased the performance in large-scale long-tail classification problems. 

This indicates that classification difficulty comes from not only data imbalance but also similarity between class samples, e.g., ``vibrato''  (majority class) and ``trill'', ``trillo'' (minority classes). These techniques exhibit both frequency and amplitude modulations, while vibrato and trill mainly rely on frequency modulation and trillo on amplitude modulation. However, close observation of trillo spectrogram also shows some frequency modulation \cite{wilkins2018vocalset}. Detecting these subtle  balance of amplitude and frequency modulations was the difficulty in this task.

\section{Conclusion}
In this paper, we proposed audio feature learning by deformable convolution and imbalance-aware learning based on classifier decoupling and a weighted inverse frequency loss, for singing technique classification. 
The experiments showed that applying deformable convolution in the last two layers and cRT with smoothed inverse frequency weights improve the classification performance. Future study can explore more complex weighting-based loss functions (e.g., \cite{cui2019class}) and evaluating our concept on real-world singing performances, in which the problems of this study (i.e., feature learning and label sparseness \cite{yamamoto2021towards}) are more serious.

\section{Acknowledgements}
This work was supported by JST SPRING, Grant Number JPMJSP2124.

\bibliography{r}
\bibliographystyle{IEEEtran}
\end{document}